\documentclass[a4paper]{jpconf}
\usepackage{graphicx}
\usepackage{amssymb, amsmath}

\newcommand{\bea}{\begin{eqnarray}}
\newcommand{\eea}{\end{eqnarray}}
\newcommand{\eq}[1]{Eq.~(\ref{#1})}

\begin{document}
\title{
{\normalsize\vskip -50pt
\mbox{} \hfill WUB/07-10 \\}
\vskip 25pt
New Higgs mechanism from the lattice
}

\author{F Knechtli$^1$, N Irges$^2$ and M Luz$^1$}

\address{$^1$ Fachbereich C, Bergische Universit{\"a}t Wuppertal,
42097 Wuppertal, Germany}
\address{$^2$ High Energy and Elementary Particle Physics Division,
Department of Physics, University of Crete, 71003 Heraklion, Greece}

\ead{knechtli@physik.uni-wuppertal.de}

\begin{abstract}
Spontaneous symmetry breaking has been observed in lattice simulations of
five-dimensional gauge theories on an orbifold. This effect is reproduced
by perturbation theory if it is modified to account for a finite cut-off.
We present a comparison of lattice and analytic results for bulk gauge
group $SU(2)$.
\end{abstract}

\section{Introduction}

Gauge theories in dimensions $d>4$ are (part of) models, which aim at 
explaining the origin of the Higgs field and electroweak symmetry breaking, one
of Gauguin's questions in particle physics \cite{Ellis:2007kd}
that has been reviewed at this conference \cite{Giudice:2007qj}.
In such extensions of the Standard Model based on gauge-Higgs unification the
Higgs is identified with extra-dimensional components of a gauge field.
The Higgs potential is zero at tree level and
is generated through quantum effects \cite{Coleman:1973jx}. The Higgs mass
and quartic coupling, which are inputs of the Standard Model, become
dynamical properties. In case of non-simply connected extra-dimensional
spaces, like a circle $S^1$ or a torus $T^2$, the physical degrees of freedom
of the Higgs actually reside in the non-contractible Polyakov loops.
It is argued that the Higgs potential is finite \cite{Antoniadis:2001cv}
despite the non-renormalizability of the models, the intuitive reason being
that non-local counterterms are not allowed. The potential
can further break spontaneously the gauge symmetry. This extra-dimensional
version of the Higgs mechanism is referred as the Hosotani mechanism
\cite{Hosotani:1983xw,Hosotani:1989bm}.

A particularly simple and attractive extra-dimensional space is the orbifold
$S^1/\mathbb{Z}_2$. The $\mathbb{Z}_2$ projection identifies degrees of
freedom under the reflection of the fifth dimensional coordinate. The circle
$S^1$ is thus projected onto an interval $I_1$.
Gauge fields are identified under this reflection up to a global gauge
transformation.
The ends of the interval are the fixed points of the reflection and naturally
define boundaries. It turns out that there is a tower of boundary conditions
for the gauge field and its derivatives, which can be derived as a limit of
a gauge invariant construction \cite{Irges:2004gy}. In this limit the gauge
invariance is broken on the boundaries. For the purpose of recovering
the Standard Model Higgs this explicit breaking of the gauge symmetry allows
to get a Higgs field which is not in the adjoint representation of the gauge
group: some of the extra-dimensional components of the gauge field are set to
zero at the boundaries and therefore do not have a zero-mode.
If we think of dimensional
reduction as in finite temperature field theory, the low-energy effective
theory is described by zero-modes and we end up with a Higgs field in the
fundamental representation of a subgroup of the original gauge group.

Five-dimensional gauge theories formulated on $S^1/\mathbb{Z}_2$ can be
studied using perturbation theory.
One starts with a Fourier or Kaluza--Klein (KK) expansion of the gauge field,
each gauge field component is associated with a tower of
four-dimensional fields but only some (even) components have a zero-mode.
The 1-loop expression for the Higgs mass $m_H$
for general gauge group $SU(N)$ is
\cite{vonGersdorff:2002as,Cheng:2002iz}
\bea
 (m_HR)^2 & = & \frac{9N\zeta(3)}{32\pi^4}g_4^2 \,,\quad
 g_4^2\,=\,\frac{g_5^2}{2\pi R} \,, \label{mh1loop}
\eea
where $R$ is the radius of the extra dimension and $g_5$, $g_4$ are the
five-dimensional and effective four-dimensional gauge couplings respectively.
This expression agrees with the one obtained from the computation of the
effective potential at 1-loop \cite{Kubo:2001zc}. This fact we find
remarkable, since the 1-loop potential is an effective potential
for free fields and the
gauge coupling there only enters indirectly through the shifted masses of the
Kaluza--Klein modes. The minimization of the potential also shows that
there is no spontaneous symmetry breaking (at 1-loop). This observation
led to consider models where (a large number of) bulk fermion fields are
added to trigger spontaneous symmetry breaking. We found it appropriate
to pause for a moment and look in detail at how the perturbative computations
have been done.

\section{Perturbative computation of the Higgs potential}

Perturbative computations of the effective scalar potential in five-dimensional
gauge theories consist of two steps: diagonalization of the mass matrix for
the KK modes and (re)summation of their 1-loop contributions to the
effective action.

The first step is done by expanding the five-dimensional fields in a Fourier
basis on $S^1/\mathbb{Z}_2$. The orbifold boundary conditions determine which
gauge field component is even and which is odd,
\bea
 E(x,x_5) & = & \frac{1}{\sqrt{2\pi R}}E^{(0)}(x) +
 \frac{1}{\sqrt{\pi R}}\sum_{n=1}^\infty
 E^{(n)}(x)\cos(nx_5/R) \quad\mbox{even fields} \,,\\
 O(x,x_5) & = & \frac{1}{\sqrt{\pi R}}\sum_{n=1}^\infty O^{(n)}(x)\sin(nx_5/R)
 \quad\mbox{odd fields} \,.
\eea
Dimensional reduction is expected to occur for energies $E\ll1/R$, where
physics is described by a low-energy effective theory of zero-modes
$E^{(0)}$. This expectation has to be verified by computations of low-energy
physical quantities. The mode expansion of the fields is inserted in the
lagrangean
\bea
-\cal L & = & \frac{1}{2g_5^2}{\rm tr}\{F_{MN}F_{MN}\}
              +\frac{1}{g_5^2\xi}{\rm tr}\{(\bar{D}_MA_M)^2
\eea
with $\bar{D}_MF = \partial_MF+[\langle A_M\rangle,F]$ and we set $\xi=1$
(unexplained notation is as in \cite{Irges:2007qq}).
From a four-dimensional point of view, the five-dimensional components
of the gauge field $A_5$ are scalars and can assume a vacuum expectation
value (vev) $\langle A_5\rangle \neq 0$. The masses of the KK modes
are found by diagonalization of the operator
$\bar{D}_5\bar{D}_5$.
In order to give a concrete example we consider the gauge group $SU(2)$ with
the orbifold breaking
\bea
 SU(2) & \stackrel{\mathbb{Z}_2}{\longrightarrow} & U(1) \label{SU2ob}
\eea
The KK masses are
\bea
 A_\mu^{3, (0)} \ (\mbox{gauge boson}):
 & (m_Z R)^2 = \alpha^2 \label{m_z}  \label{SU2gb0} \\
 A_5^{1, 2 (0)} \ (\mbox{Higgs}):
 & (m_{A_5} R)^2 = \alpha^2\,,\; 0 \label{m_h} \label{SU2h0} \\
 \mbox{higher KK modes}: & (m_nR)^2 =n^2\,,\; (n\pm \alpha)^2
 \label{SU2kkn}
\eea
where
\bea
 \alpha \,=\, g_5\langle A_5^1\rangle R \,=\,
 g_5 v R/\sqrt{2\pi R}
\eea
is a dimensionless modulus parametrizing the four-dimensional Higgs vev $v$.

The second step is to sum the four-dimensional 1-loop effective action for
the KK modes. This step involves a Poisson resummation which eliminates a
constant (i.e. independent of $\alpha$) divergent contribution. The result is
the periodic potential
\bea
 V & = & -\frac{9}{64\pi^6R^4}\sum_{m=1}^{\infty}\frac{\cos(2\pi m\alpha)}{m^5}
 \,,
\eea
which has degenerate minima at
$\alpha = \alpha_{\rm min} = 0\; {\rm mod}\, \mathbb{Z}$. For these values
of $\alpha$ the spectrum as a whole is the same as for $\alpha=0$. There is no
spontaneous symmetry breaking of the remnant $U(1)$ gauge symmetry in
\eq{SU2ob}, which would manifest itself in a massive lowest mode for $A_\mu^3$.

\section{Lattice simulations of gauge group $SU(2)$}

\begin{figure}[h]
\begin{minipage}{18pc}
\includegraphics[width=18pc]{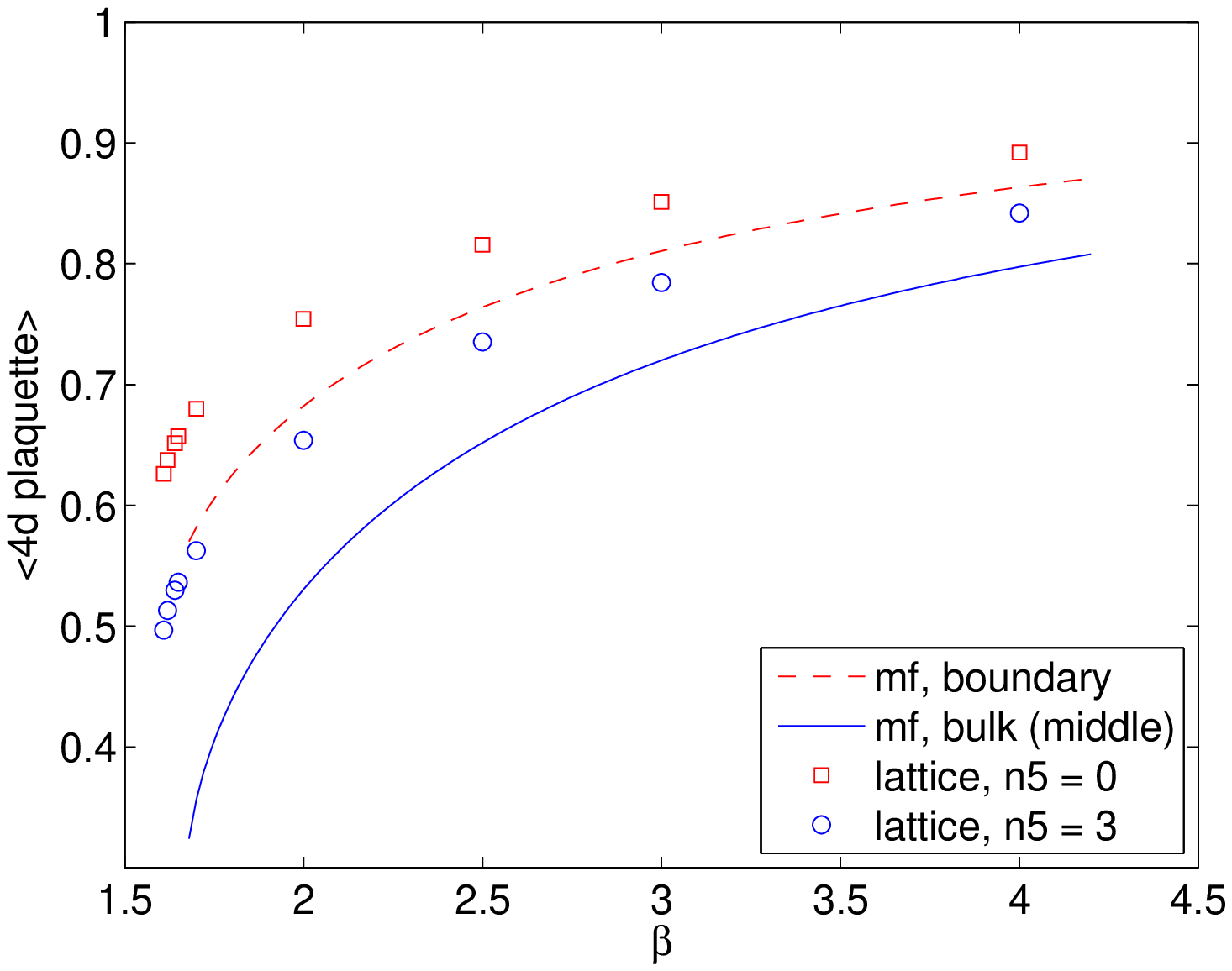}
\caption{\label{f_4d_plq_mf}
Average plaquette in four-dimensional hyperplanes in the boundary
(squares) and in the bulk (circles) of the orbifold with
$T/a = 96$, $L/a = 14$, $N_5=6$. Comparison to the mean-field
calculation (lines).}
\end{minipage}\hspace{2pc}%
\begin{minipage}{18pc}
\includegraphics[width=18pc]{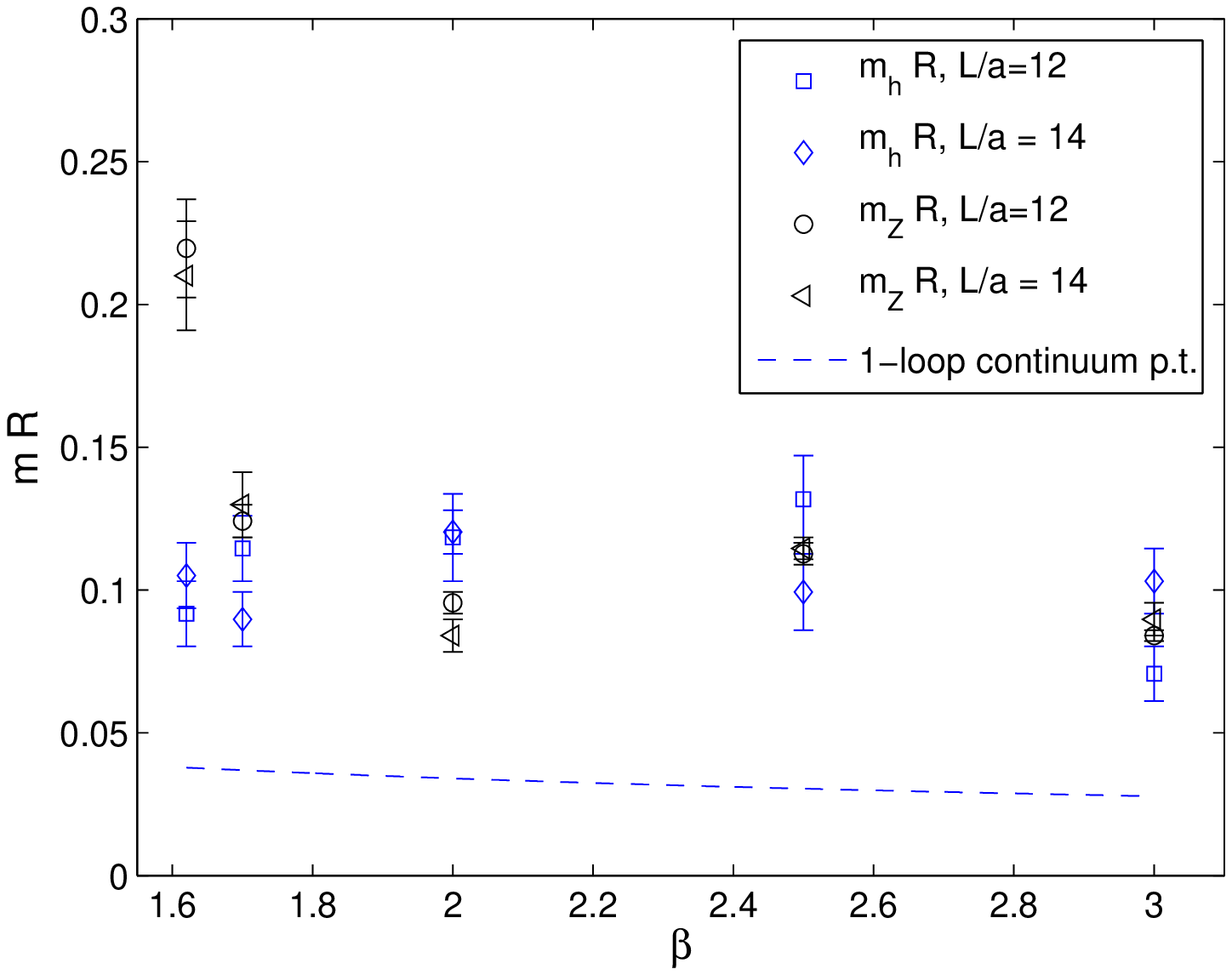}
\caption{\label{f_spectrum}
Spectrum from lattice simulations of the orbifold with
$T/a = 96$, $L/a = 12$ and $L/a=14$, $N_5=6$. The dashed
line represents \eq{mh1loop}.}
\end{minipage}
\end{figure}

The orbifold theory with the explicit symmetry breaking \eq{SU2ob} can be
defined on a Euclidean space-time lattice \cite{Irges:2004gy}. The
geometry is the strip $\{z=a(n_\mu,n_5)|\;0\le n_5\le N_5\}$, where
$a$ is the lattice spacing, the integer coordinates $n_\mu$ label
points in a four-dimensional hypercube $\frac{T}{a}\times(\frac{L}{a})^3$ and
$n_5=0,\;N_5$ define the boundaries.
The parameter space on the lattice is given by
\bea
 N_5\;=\;\pi R\Lambda \quad\mbox{and}\quad \beta=2N/(g_5^2\Lambda) \,,
\eea
where $\Lambda=1/a$ is the ultraviolet cut-off. Details on the lattice
action and operators can be found in \cite{Irges:2006hg}. It turns out
that there is a first order phase transition at $\beta_c\simeq1.6$.
This transition is the same as the one observed in infinite volume
(or with periodic boundary conditions)
\cite{Creutz:1979dw,Beard:1997ic,Ejiri:2000fc,Farakos:2002zb} and
can be detected by a jump in the expectation value of the plaquettes.
We did a mean-field calculation \cite{Knechtli:2005dw} that also shows
the presence of the phase transition and reproduces the qualitative
behavior of the plaquettes, see figure \ref{f_4d_plq_mf}. The spectrum
of the scalar (Higgs/glueball) and vector (gauge boson) states can only
be measured in simulations for $\beta>\beta_c$ and is shown in
figure \ref{f_spectrum} for $N_5=6$ as a function of $\beta$. The Higgs
mass is larger then the 1-loop continuum value \eq{mh1loop}. The gauge
boson is a massive $Z$ boson, contrary to the perturbative result that
we discussed in the previous section. This is the first lattice evidence for
spontaneous symmetry breaking in pure extra-dimensional gauge theories
\cite{Irges:2006zf}. The appearance of a Higgs phase is not completely
unexpected from dimensional reduction \cite{Fradkin:1978dv}.

\section{Perturbative computations with a cut-off}

\begin{figure}[h]
\includegraphics[width=18pc]{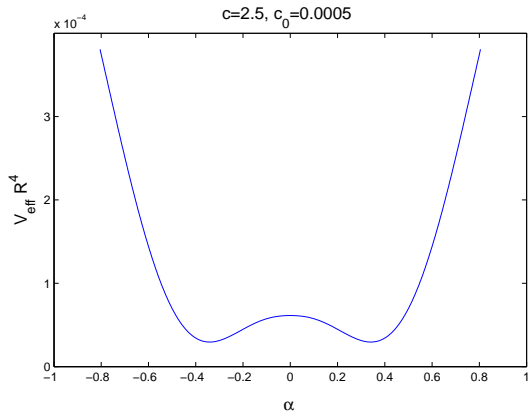}\hspace{2pc}%
\begin{minipage}[b]{18pc}
\caption{\label{f_su2_pot}
Perturbative Higgs potential for $N_5=6$ with cut-off effects.}
\end{minipage}
\end{figure}

Five-dimensional gauge theories are trivial: if a ultraviolet cut-off
$\Lambda$ is removed from the theory, the four-dimensional effective
coupling $g_4$ goes to zero \cite{Dienes:1998vg}, see also
\cite{Gies:2003ic,Morris:2004mg}. In this limit the Higgs mass \eq{mh1loop}
tends to zero. In order to move away from the trivial limit we
regularize the theory on a Euclidean lattice, which naturally
provides a gauge invariant cut-off $\Lambda=1/a$.
We make the hypothesis that in a vicinity of the trivial point the 
lattice theory
can be described by a continuum Symanzik effective lagrangean
\bea
-{\cal L} & = & \frac{1}{2g_5^2}{\rm tr}\{F_{MN}F_{MN}\}+
\sum_{p_i}{c^{(p_i)}(N_5,\beta)}\;{a^{p_i-4}}\;{\cal O}^{(p_i)}+\ldots \,,
\label{sym}
\eea
where ${\cal O}^{(p_i)}$ is an operator of dimension $p_i>4$.
This expansion has been shown to describe cut-off effects for
renormalizable theories
\cite{Symanzik:1981hc,Symanzik:1983dc,Symanzik:1983gh,Luscher:1998pe}.
Our working hypothesis is that the five-dimensional gauge theory, despite its
non-renormalizability, possesses a scaling regime where it is described
by \eq{sym}.

For the orbifold the operators of lowest dimension are
\bea
 c_0\,{\cal O}^{(5)} & = & \frac{\pi a{c}_0}{4}
 F_{5\mu}^{\hat a}F_{5\mu}^{\hat a}
 \left[\delta(x_5) + \delta(x_5-\pi R)\right] \,, \label{O5} \\
 c\,{\cal O}^{(6)} & = &
\sum_{M,N} \frac{c}{2}{\rm tr}\{F_{MN}(D_M^2+D_N^2)F_{MN}\} \,, \label{O6}
\eea
where the coefficients $c_0\equiv c^{(5)}(N_5,\beta)$ and
$c\equiv c^{(6)}(N_5,\beta)$ depend on the lattice gauge action and
can be computed in perturbation theory.
For example $c=1/12$ at tree level for the Wilson plaquette action.
Higher derivative operators like \eq{O6} appear in models for
new Higgs physics considered in \cite{Grinstein:2007mp,Fodor:2007fn},
where they are interpreted as new particles (ghosts) which cancel
quadratic divergences in the Higgs mass.

The operators \eq{O5} and \eq{O6} induce corrections to the KK masses
of the gauge field $A_\mu$. For example \eq{SU2gb0} and \eq{SU2kkn}
in the $SU(2)$ model are changed to
\bea
(m_nR)^2 =
&& n^2\, , \qquad\mbox{for}\;n>0 \,,\\
&& (n\pm \alpha)^2 + \frac{c_0\alpha^2}{2}\frac{\pi}{N_5}+
c\, (n\pm \alpha)^4\frac{\pi^2}{N_5^2} \qquad\mbox{for}\;n\ge0 \,,
\label{gbmasses}
\eea
where we keep only the leading $n$-independent correction from
the boundary term. The gauge field $A_5$ and the ghost field do
not receive cut-off corrections, since the masses of their KK modes
originate from the gauge fixing term.

When inserted into the formula for the Higgs potential, the corrected
KK masses \eq{gbmasses} lead to an expansion in $c_0$ and $c$ of the $A_\mu$
(gauge) contribution \cite{Irges:2007qq}. The latter is defined as
\bea
V^{\rm gauge} & = &  - \frac{1}{2} \sum_{n\in\mathbb{Z}}
\int \limits_0^\infty \frac{{\rm d}l}{l}
{\rm e}^{-\frac{1}{l}(m_n^2a^2 + 8)}\frac{1}{a^4}{\rm I}_0^4
\left(\frac{2}{l}\right) \,. \label{Vgauge}
\eea
A tricky point here is that we extend the summation over a finite number of
KK modes on the lattice to $n\to\infty$, but this is justified since the
contribution of higher modes
is exponentially suppressed in \eq{Vgauge}. The expansion then reads
\bea
 V^{\rm gauge} & = & f_0 + \underbrace{c_0f_1}_{{\rm O}(a)}
  + \underbrace{c_0^2\tilde{f}_2 + cf_2}_{{\rm O}(a^2)}
\eea
and the total potential is given by
\bea
V & = & \underbrace{ 4V^{\rm gauge}}_{A_\mu}
        \underbrace{+V^{\rm scalar}}_{A_5}
        \underbrace{-2V^{\rm scalar}}_{\mbox{ghosts}} \,.
\eea
Here $V^{\rm scalar}$ is obtained by setting $c=c_0=0$ in \eq{Vgauge}.
The cut-off corrected effective potential for $N_5=6$,
$c=2.5$ and $c_0=0.0005$ is shown in figure \ref{f_su2_pot}.
For each value of $N_5$ and $c_0=0$,
there is a minimal positive value of $c$ such that there is
spontaneous symmetry breaking \cite{Luz:2007vg}:
the minimum of the potential is attained at $\alpha_{\rm min}=1/2$.
When in addition $c_0>0$,
the potential is not any more periodic in $\alpha$. The
requirement that the vev is below the cut-off scale: $v<1/a$ implies
the constraint $|\alpha| < \sqrt{N N_5/(\pi^2 \beta)}$.
The Higgs potential has the characteristic shape like in the
Standard Model as shown figure \ref{f_su2_pot} and the minimum
is shifted continuously in the range
$0<\alpha_{\rm min}<1$ depending on the value of $c_0$.

\section{Comparing perturbation theory with lattice results}

\begin{figure}[h]
\begin{minipage}{18pc}
\includegraphics[width=18pc]{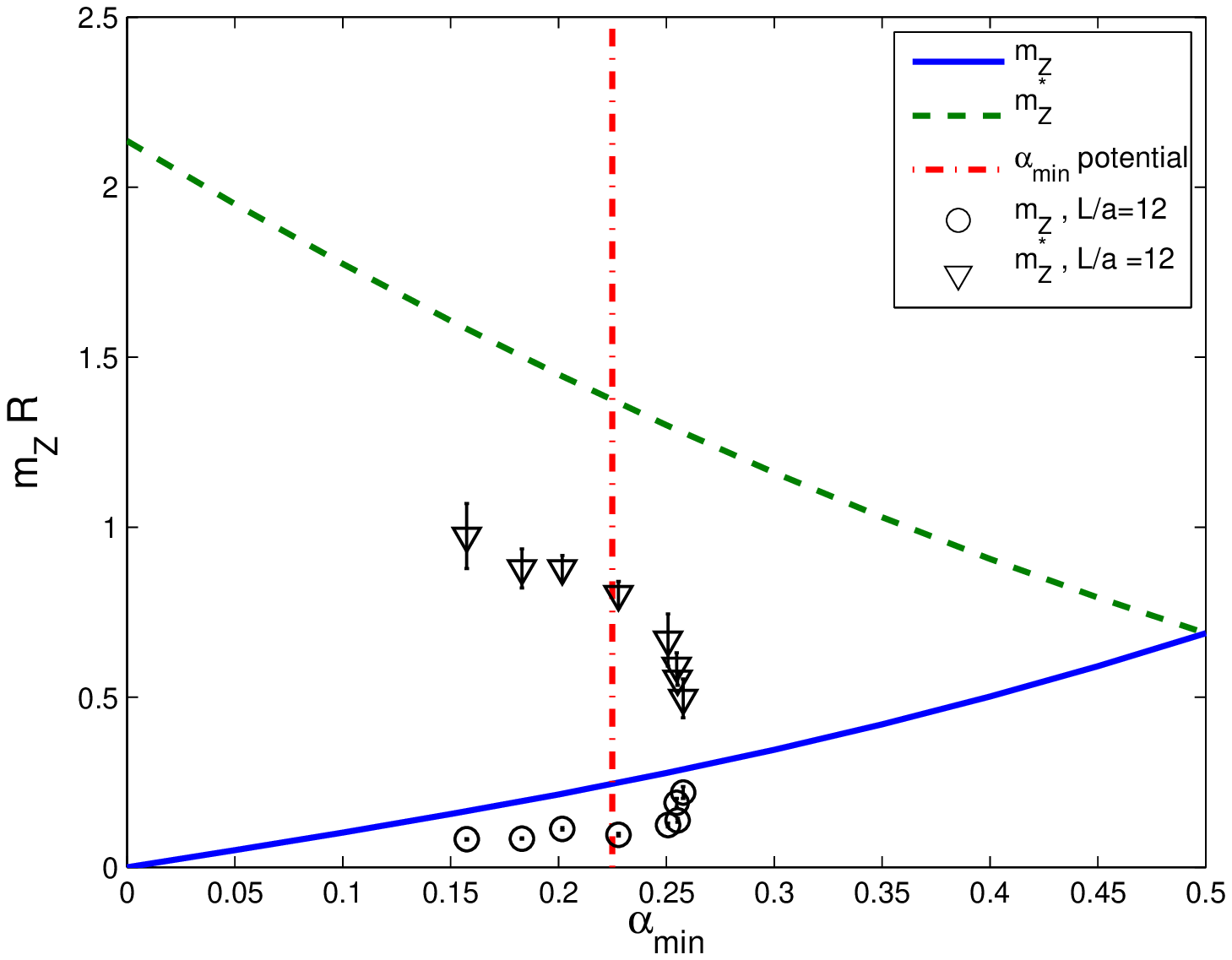}
\caption{\label{f_mz_alpha}
Comparison of the lattice masses for the $Z$ and $Z^*$ bosons with
the Kaluza--Klein masses, which are corrected to include cut-off effects.}
\end{minipage}\hspace{2pc}%
\begin{minipage}{18pc}
\includegraphics[width=18pc]{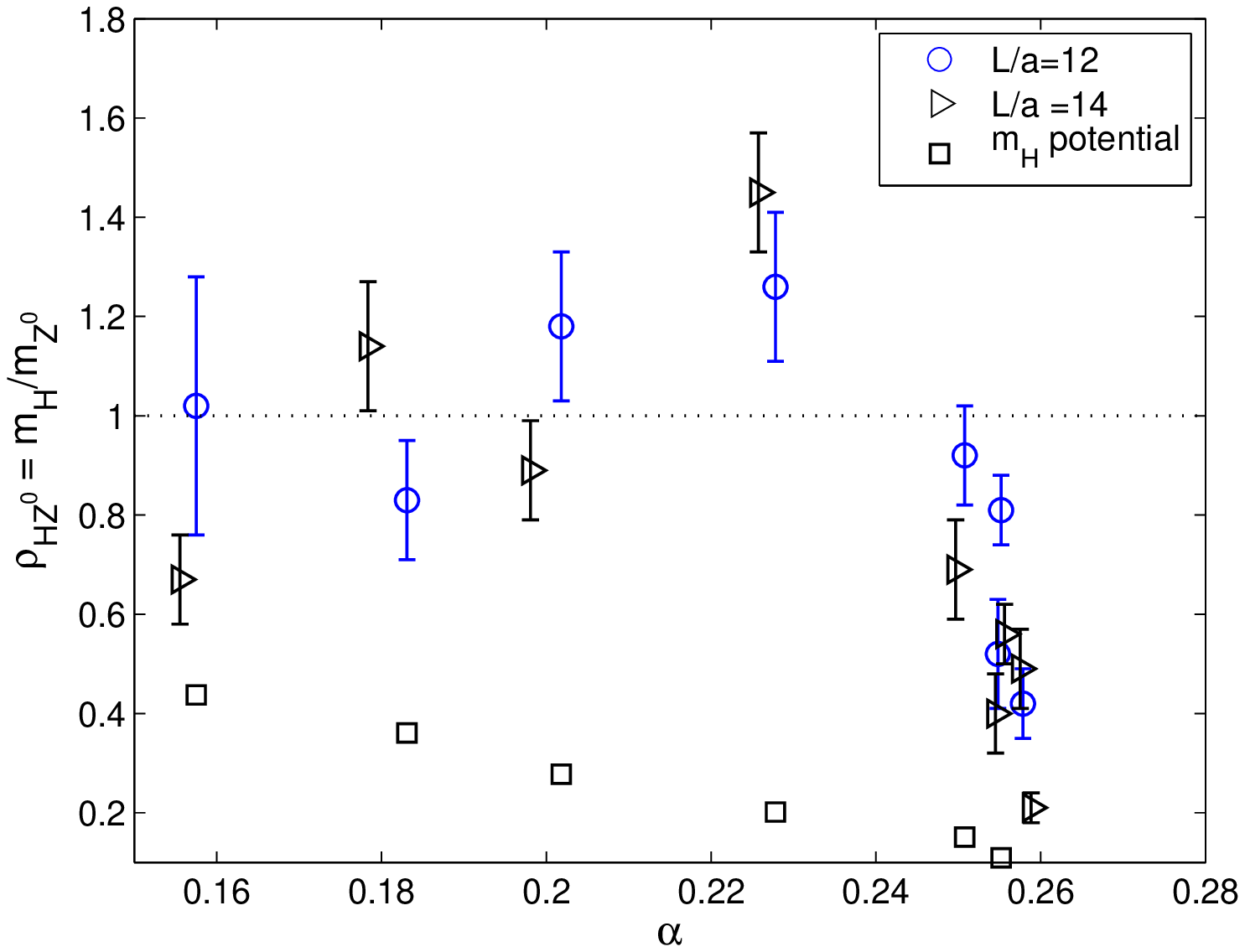}
\caption{\label{f_rho}
Ratio of the Higgs to the $Z$-boson mass, from lattice simulations
and analytic computations.}
\end{minipage}
\end{figure}

We are now in the position to compare the results from cut-off corrected
perturbation theory with the lattice simulations of the orbifold with
bulk gauge group $SU(2)$. This should provide the justification for our
working hypothesis of using the Symanzik expansion. In figure \ref{f_mz_alpha}
we compare for $N_5=6$ the masses of the $Z$ and $Z^*$ bosons as a
function of the modulus $\alpha_{\rm min}$. The lines represent
the formulae \eq{gbmasses} for $n=0,1$ for fixed values $c=13.0$ and 
$c_0=0.0121$ varying $\alpha=\alpha_{\rm min}$.
Actually for these values of $c$ and $c_0$ the potential has
a minimum at the position indicated by the vertical dotted line.
The symbols are the simulation results of the orbifold with
$T/a = 96$, $L/a = 12$, $N_5=6$, plotted using
a lattice determination of $\alpha_{\rm min}$ \cite{Irges:2007qq}. There is
good qualitative agreement.

In figure \ref{f_rho} we compare always for $N_5=6$ the ratio of the Higgs
to the Z-boson mass $\rho_{HZ^0} = m_H/m_{Z^0}$.
Here $c$ and $c_0$ in the potential
calculation are tuned to give the minimum at the value $\alpha_{\rm min}$ as
it is determined in the lattice simulation. The perturbative results for
$\rho_{HZ^0}$ are represented by the square symbols. The lattice results
(circles and triangles, orbifold with
$T/a = 96$, $L/a = 12$ and $L/a=14$, $N_5=6$) indicate that
contrary to perturbation theory on the lattice it is possible to get
$\rho_{HZ^0}\ge1$.

\section{Conclusions}
We are investigating five-dimensional gauge theories
as models to derive electroweak symmetry breaking, by a combination
of lattice simulations and analytic computations. We gave the first
evidence for spontaneous symmetry breaking and could reproduce this
effect by the inclusion of cut-off effects in perturbation theory.
Our results encourage to pursue the study in order to establish whether
these theories possesses a scaling regime, where the values of
physical observables do not strongly depend on the cut-off.

\section*{References}
\bibliography{extrahiggs}           
\bibliographystyle{iopart-num}

\end{document}